\documentclass[a4paper,11pt]{article}

\usepackage{amssymb}
\usepackage{amsmath}
\usepackage{graphicx}
\usepackage{comment}
\usepackage{fancyhdr}

\usepackage{geometry}
\begin{document}

\title{Benford's law and complex atomic spectra}

\author{Jean-Christophe Pain\footnote{CEA/DIF, B.P. 12, 91680 Bruy\`eres-Le-Ch\^atel Cedex, France, email: jean-christophe.pain@cea.fr}}

\maketitle

\begin{abstract}
We found that in transition arrays of complex atomic spectra, the strengths of electric-dipolar lines obey Benford's law, which means that their significant digits follow a logarithmic distribution favoring the smallest values. This indicates that atomic processes result from the superposition of uncorrelated probability laws and that the occurrence of digits reflects the constraints induced by the selection rules. Furthermore, Benford's law can be a useful test of theoretical spectroscopic models. Its applicability to the statistics of electric-dipolar lines can be understood in the framework of Random Matrix Theory and is consistent with the Porter-Thomas law.
\end{abstract}

\section{Introduction}

Newcomb \cite{new} noticed that the beginning pages of logarithm books were more used than the last pages. This observation led to the conjecture that the significant digits of many sets of naturally occurring data describing the physical world are not equi-probably distributed, but in a way that favors smaller significant digits. For instance, the first significant digit (\emph{i.e.} the first digit which is non zero) will be 5 more frequently than 6 and the first three significant digits will be 458 more often than 462. Moreover, this result is scale invariant. Benford provided a probability distribution function for significant digits. The probability that the $N$ first significant digits $d_i$ ($i=1,N$) are equal respectively to $k_i$
($k_i$=1,9) is given by \cite{ben}:
\begin{equation}\label{bl1}
P(d_i=k_i)=\log_{10}\Big(1+\Big[\sum_{i=1}^Nd_i\times10^{N-i}\Big]^{-1}\Big).
\end{equation}
For instance, since some accounting data satisfy this rule, it is used to detect fraud \cite{nig}. River lengths, mountain heights, populations of cities \cite{tor}, radioactive decay half lives \cite{buc}, physical constants and some mathematical series satisfy Benford's law. However, if many data roughly satisfy the law, perfect agreement is not the general case and the law is not universal. In the present work, focus will be put on the first significant digit, which probability of being equal to $k$ is given by 
\begin{equation}\label{bl2}
P(d_1=k)=\log_{10}(1+k^{-1}). 
\end{equation}

\section{Digits of line strengths}

\begin{figure}
\begin{center}
\includegraphics[width=8.6cm]{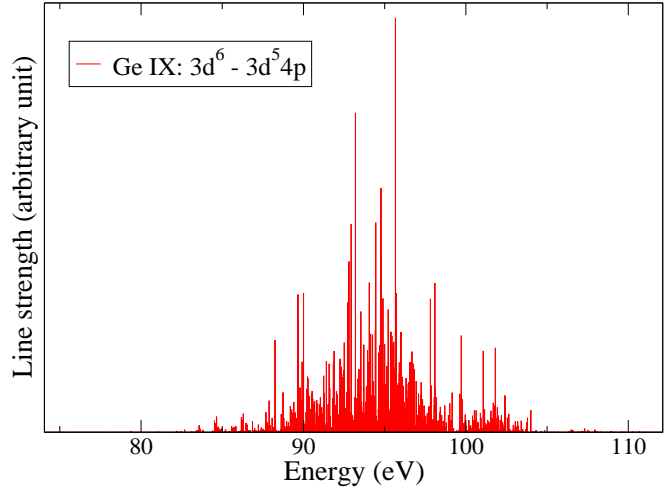}
\end{center}
\caption{Array of lines for the transition $3d^6 \rightarrow 3d^54p$ of Ge IX.}
\label{fig1}
\end{figure}

\begin{figure}
\begin{center}
\includegraphics[width=8.6cm]{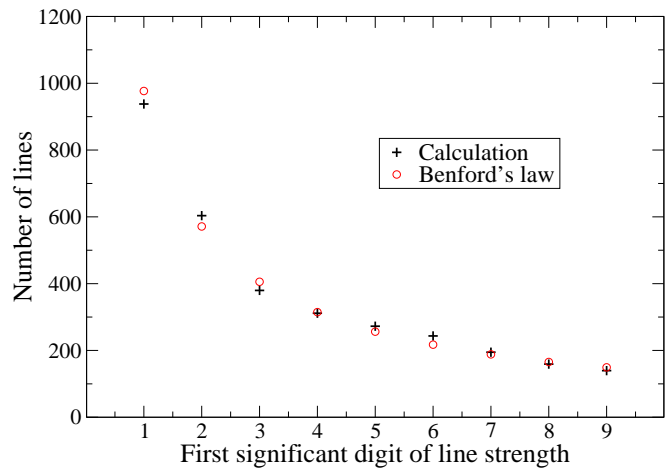}
\end{center}
\caption{Fraction of lines versus first significant digit for the transition
$3d^6 \rightarrow 3d^54p$ of Ge IX. The total number of lines is 3245.}
\label{fig2}
\end{figure}

To our knowledge, although a great attention has been paid to the modeling of electric-dipolar (E1) transition lines in a
complex atomic spectrum typical of a hot plasma, it has never been checked whether the statistics of the lines obeys Benford's law. Most of the studies focused on the distribution of the values of the line strengths (see for instance \cite{bau5} or \cite{lea}), but not on the significant digits. Figure \ref{fig1} represents the transition array $3d^6 \rightarrow 3d^54p$ of Ge IX calculated with Cowan's \textit{ab initio} atomic-structure code \cite{cow}, which provides the Slater direct and exchange integrals, and the one-particle dipole-moment integrals evaluated from Hartree-Fock wavefunctions. Figure \ref{fig2} displays the number of lines of that transition array for each value of the first significant digit calculated from our approach and predicted from Benford's law. The total number of lines is 3245. It is clear that the two distribution of lines are very similar. The same property has been observed on other transition arrays for different elements. Table \ref{tab1} displays the fraction of lines per significant digit for four transitions: $3d^6 \rightarrow 3d^54p$ (3245 lines) for Ge IX ($T_1$), $3d^4 \rightarrow 3d^34f$ (2825 lines) for Ta LII ($T_2$), $3d^24s4p \rightarrow 3d4s4p^2$ (2722 lines) for Br XIV ($T_3$) and $3d^24s4p^3 \rightarrow 3d4s4p^4$ (8231 lines) for Br XII ($T_4$). It is worth mentioning that even for transition arrays with a much smaller number of lines, the law still applies very well.

\section{Testing spectroscopic models}

In a hot plasma, the total spectrum consists of a huge number of transition arrays. Therefore, the total number of lines can be immense. Usually, when the lines coalesce due to the physical broadening processes (natural width, electron impact, Stark effect and Doppler effect), the distribution of lines in intermediate coupling is modeled by a continuous (Gaussian) distribution, named UTA (Unresolved Transition Array) which average energy and variance are evaluated following the work of Bauche \emph{et al.} \cite{bau1}. However, in that case some information is lost, and random calculations can help ``re-constructing'' approximately the resolved distribution of lines ensuring known physical rules \cite{fan}. 

The RTA (Resolved Transition Arrays) approach developed by Bauche \emph{et al.} \cite{bau2} and recently improved by Gilleron \emph{et al.} \cite{gil} enables one to simulate E1 lines of a transition array avoiding direct diagonalization of the atomic Hamiltonian. The strength of a line and the energies of the lower and upper levels are picked up at random satisfying some specific constraints: preservation of the number of lines, of the total strength and weighted variance of the transition array, as well as of the unweighted variances of levels of lower and upper configurations. In that approach, the distribution function also includes a correlation between the energies and amplitudes of the lines (propensity law). Therefore, Benford's significant-digit law can be a performing test of the validity of the correlation rules included in such approaches.

\begin{table}
\begin{center}
\begin{tabular}{cccccc}\hline
$k$ & $T_1$ & $T_2$ & $T_3$ & $T_4$ & Benford\\\hline\hline
1 & 0.289 & 0.307 & 0.307 & 0.300 & 0.301\\\hline
2 & 0.186 & 0.182 & 0.179 & 0.186 & 0.176\\\hline
3 & 0.117 & 0.121 & 0.126 & 0.117 & 0.125\\\hline
4 & 0.096 & 0.096 & 0.093 & 0.097 & 0.097\\\hline
5 & 0.084 & 0.079 & 0.079 & 0.076 & 0.079\\\hline
6 & 0.075 & 0.064 & 0.071 & 0.071 & 0.067\\\hline
7 & 0.060 & 0.060 & 0.055 & 0.055 & 0.058\\\hline
8 & 0.049 & 0.051 & 0.049 & 0.050 & 0.051\\\hline
9 & 0.043 & 0.039 & 0.043 & 0.048 & 0.046\\\hline
\end{tabular}
\caption{Fraction of lines which first digit is $k$ for transitions $T_1$, $T_2$, $T_3$ and $T_4$ compared to the values predicted by Benford's law.}\label{tab1}
\end{center}
\end{table}

\section{Interpretation}

The fact that E1 lines satisfy Benford's law indicates that the distribution of digits reflects the symmetry due to the selection rules. If transitions were governed by uncorrelated random processes, each digit would be equi-probable. Benford's law is still not fully understood mathematically. However, some key elements are known since the middle of the nineties. For instance, Benford's law holds for any rescaling of the data, since it does not depend on any particular choice of units \cite{rai}. It is the only scale invariant law referring to digits since the logarithmic distribution is the unique continuous base-invariant distribution \cite{hil2}. More precisely, if probability distributions are selected at random and random samples are taken from each of these distributions so that the overall process is base invariant, then the significant-digit frequencies of the sample will converge to the logarithmic distribution. In other words, the more  diversified the probability distributions are, the better the data sets fit Benford's law. The understanding of why systems with many interacting particles spontaneously organize into scale-invariant states is a difficult task of statistical physics. 

Moreover, Benford's law applies if the system is governed by multiplicative processes \cite{pie}. Indeed, a random multiplicative process corresponds to an additive process (dynamical description of a Brownian process) in a logarithmic space. In Wigner's Random Matrix Theory \cite{meh,cam}, the Hamiltonian is defined in the Gaussian Orthogonal Ensemble (GOE) by an ensemble of real symmetric matrices which probability distribution is a product of the distributions for the individual matrix elements  $H_{kl}$, which are considered as stochastic variables. The variance of the distribution for the diagonal elements is twice the one for the off-diagonal elements. The line strength $S$ is proportional to $\langle i|\vec{D}|j\rangle|^2$, where $\vec{D}$ is the dipolar operator and $|i\rangle$ and $|j\rangle$ are eigenvectors of the Hamiltonian. Therefore, the line strengths involve quantities such as products of $H_{kl}$ and one has
\begin{equation}\label{sto}
\frac{S}{S'}=\zeta,
\end{equation}
where $S$ and $S'$ are the strengths of two lines belonging to the transition array and $\zeta$ is a stochastic variable. Eq. \ref{sto}) can be written
\begin{equation}
\log S=\log\zeta+\log S'
\end{equation}
The central-limit theorem states that the probability distribution that the value of the $n^{th}$ strength is $S$ will be Gaussian with a variance $\propto n^{1/2}$. In the infinite limit, the distribution will approach the uniform one, characterized by constant $K$. Therefore, one has
\begin{equation}
\int P(\log S)d(\log S)=K\int\frac{dS}{S}.
\end{equation}
The probability $P$ that the first significant digit $d_1$ of $S$ is $k$ in base 10 is given by
\begin{equation}
P(d_1=k)=\int_{k}^{k+1}\frac{dS}{S}\Big/\int_{1}^{10}
\frac{dS}{S}=\log_{10}\Big(1+\frac{1}{k}\Big),
\end{equation}
which is exactly Eq. (\ref{bl2}). This heuristic argument relies on the idea that fluctuations are governed by multiplicative processes involving a stochastic variable. The matrix elements of the Hamiltonian are correlated stochastic variables and the product of such variables leads to Benford's logarithmic distribution of digits.

Another argument can be invoked to understand the mathematical fundation of Benford's law. Porter and Thomas have shown that the amplitudes of the lines between all the levels of two random matrices obey a Gaussian distribution \cite{por1,por2}. The
strength being the square of the amplitude, its distribution is \cite{bau5,bau6}
\begin{equation}\label{port}
D(S)=\frac{L}{\sqrt{2\pi <S>S}}\exp[-\frac{S}{2<S>}],
\end{equation}
where $L$ and $<S>$ are respectively the number of lines and the average value of the line strength in a ($J,J'$) set. When numbers are taken from an exponential distribution, they automatically obey Benford's law. Therefore, when the exponential term dominates in $D(S)$, which is often the case except close to the origin (\emph{i.e.} for very weak lines), Benford's law applies.

The Random Matrix Theory contains approximate symmetries, which are not sufficient to describe the vicinity of Russell-Saunders and $jj$ couplings \cite{bau3,bau4}. In the model of Wilson \emph{et al.} \cite{wil}, diagonal terms are calculated in a pure coupling using Cowan's code mentioned above and off-diagonal elements are populated statistically beyond the GOE according to a bi-Gaussian distribution function where elements are correlated. They observed a disproportionally large number of off-diagonal elements of small amplitude. However, we found that even when the Random Matrix Theory is expected to be inappropriate, \emph{i.e.} close to a pure coupling, the line strengths still fit Benford's law. This is due to the fact that even if the number of weak emerging lines is important (when the term $\frac{1}{\sqrt{S}}$ dominates in the Porter-Thomas law (\ref{port})), the number of decades is sufficiently large so that the weight of those weak lines is negligible in the statistical occurrence of digits.
 
\section{Conclusions}

The results of this study are following. First, the distribution of lines in a given transition array follows very well Benford's logarithmic law of significant digits, which was, to our knowledge, never observed and rather unexpected. 

Second, this indicates that the correlations due to the selection rules manifest themselves in the distribution of digits, even in intermediate coupling, and that such a symmetry results from a superposition of different probability laws. 

Third, computational methods avoiding diagonalization such as the Resolved Transition Arrays model, which ensures known physical rules (propensity law, correlation laws), can be improved and/or tested by imposing the constraint that the statistics of the lines must verify Benford's law with a high accuracy. 

Finally, since Benford's law can be explained in terms of a dynamics governed by multiplicative stochastic processes, the Random Matrix Theory is probably an interesting pathway for the calculation of large atomic-dipole transition arrays and Benford's law can help clarifying the existence of different classes of stochastic Gaussian variables. Moreover, Benford's law is a signature of the Porter-Thomas distribution of lines. 

%===============================================================================
%	ACKNOWLEDGMENTS
%===============================================================================

\section{Acknowledgments}

The author would like to thank F. Gilleron for numerous helpful discussions.

%===============================================================================
%	BIBLIOGRAPHY
%===============================================================================

\clearpage

\end{document}